\documentclass{article}
\usepackage{amsmath,graphicx}

\usepackage{amsmath,graphicx}
\usepackage{algorithm}
\usepackage{algpseudocode}
\usepackage{pifont}
\newcommand{\xmark}{\ding{53}}%
\usepackage{amssymb}
\usepackage{bm}
\usepackage{multirow}
\usepackage{adjustbox}
\usepackage{tabularx}
\usepackage{tikz}
\usepackage{caption}
\usepackage{subcaption}
\usepackage{verbatim}
\usepackage{float}
\usepackage{commath}
\usepackage{url}
\usepackage{color, colortbl}

\usepackage[backend=bibtex,style=ieee,minbibnames=1,maxbibnames=4]{biblatex}
\usepackage{spconf}
\usepackage[keeplastbox]{flushend}
\definecolor{LightCyan}{rgb}{0.88,1,1}

\title{Multichannel Speech Enhancement without Beamforming}
\name{Ashutosh Pandey$^\textup{1*}$, Buye Xu$^\textup{1}$, Anurag Kumar$^\textup{1}$, Jacob Donley$^\textup{1}$, Paul Calamia$^\textup{1}$ and DeLiang Wang$^\textup{2}$\thanks{$^{*}$Work done during internship at Facebook Reality Labs Research.}}
\address{$^\textup{1}$Facebook Reality Labs Research, USA\\
$^\textup{2}$Department of Computer Science and Engineering, The Ohio State University, USA}
\begin{document}
\maketitle
\ninept
\begin{abstract}
\begin{comment}
Supervised learning using deep neural networks (DNNs) is the current mainstream for multi-channel speech enhancement.
\end{comment}
Deep neural networks are often coupled with traditional spatial filters, such as MVDR beamformers for effectively exploiting spatial information. Even though single-stage end-to-end supervised models can obtain impressive enhancement, combining them with a traditional beamformer and a DNN-based post-filter in a multistage processing provides additional improvements. In this work, we propose a two-stage strategy for multi-channel speech enhancement that does not require a traditional beamformer for additional performance. First, we propose a novel attentive dense convolutional network (ADCN) for estimating real and imaginary parts of complex spectrogram. ADCN obtains state-of-the-art results among single-stage models. Next, we use ADCN with a recently proposed triple-path attentive recurrent network (TPARN) for estimating waveform samples. The proposed strategy uses two insights; first, using different approaches in two stages; and  second, using a stronger model in the first stage. We illustrate the efficacy of our strategy by evaluating multiple models in a two-stage approach with and without a traditional beamformer.
\end{abstract}
\section{Introduction}
\label{sec:intro}
Multi-channel speech enhancement is the task of removing noise, intereference and reverberation from a degraded speech signal by utilizing recordings from multiple microphones. Traditional approaches use linear spatial filters, such as those from a minimum-variance distortionless-response (MVDR) optimization, to preserve signal from the target source and suppress all other signals in the space~\cite{benesty2008microphone}. In recent years, supervised speech enhancement using deep neural  networks (DNNs) has become the mainstream methodology for speech enhancement~\cite{wang2017supervised}.  

For multi-channel processing, DNNs are generally incorporated with traditional spatial filters ~\cite{erdogan2016improved, heymann2016neural, gannot2017consolidated}, where the role of DNN is to provide better estimates of speech and noise statistics for the spatial filter. Another general approach is to train DNNs with spatial features, such as inter-channel phase, time, and level differences~\cite{wang2018combining, wang2018multi}. A DNN trained with spatial features is expected to exploit spatial cues for improved discrimination between target and interference. In recent times, end-to-end supervised approaches without any explicit spatial filtering have obtained impressive results~\cite{wang2020multi,  luo2020end, tolooshams2020channel, tzirakis2021multi, pandey2022tparn, pandey2022tadrn}. The goal of end-to-end approaches is to make the spatial filtering an implicit part of supervised learning.  

Even though these end-to-end supervised approaches have  shown impressive enhancement performance, they are yet to be widely accepted. This is due to a confounding empirical finding that an end-to-end supervised model when combined with a traditional spatial filter, such as an MVDR beamformer, and a DNN-based post-filter, provides superior results compared to a DNN-only single-stage or multistage processing ~\cite{wang2020deep, wang2020multi, wang2020complex, wang2021multi}. For example, study in ~\cite{wang2020multi} obtained impressive performance by training a dense convolutional recurrent network (DCRN) for multi-channel complex spectral mapping. However, the performance was further improved by using an MVDR beamformer along with a following DCRN as the post-filter. 

The effectiveness of an MVDR beamformer even with strong DNN models can be attributed to the fact that DNNs introduce nonlinear distortions in the enhanced speech, which are removed or reduced when they are combined with a distortionless beamformer. As a result, many of the approaches based on end-to-end learning have been inspired from traditional beamformers ~\cite{wang2018all, zhang2021adl}. The computation of beamformer weights requires matrix inversion, which makes end-to-end learning unstable. A widely accepted strategy to avoid training instability is diagonal loading ~\cite{mestre2003diagonal}. A recent study used recurrent neural networks for directly estimating the matrix inverse ~\cite{zhang2021adl}.   

In this work, we argue that the use of a traditional beamformer with a DNN is not necessary to obtain distortionless speech enhancement. We propose a novel two-stage approach where both stages are based on neural networks. Our two-stage scheme for multi-channel speech enhancement uses two key strategies. The first strategy is to use  two different approaches for speech enhancement in two stages. For instance, one stage might rely on complex spectral mapping, an approach that estimates real and the imaginary parts of complex spectrogram, and the other stage may use waveform mapping where direct waveform to waveform enhancement is done. We believe that complex spectral mapping and waveform mapping complement each other in terms of removing the overall model bias. In other words, they can get rid of some component of each other’s distortions, and hence provide an overall system with fewer distortions.The second strategy is to use the stronger approach out of the two, in the first stage of the two-stage processing. 

To this end, we first propose a novel attentive dense convolutional network (ADCN) for multi-channel complex spectral mapping. Similar to a waveform mapping based model in~\cite{pandey2021dense}, ADCN is an encoder-decoder based UNet architecture where layers within the encoder and decoder are augmented with dense blocks and attention blocks for context aggregation. ADCN obtains state-of-the art results among single-stage systems.
\begin{figure*}[!t]
\centering
\includegraphics[width=0.96\textwidth, keepaspectratio]{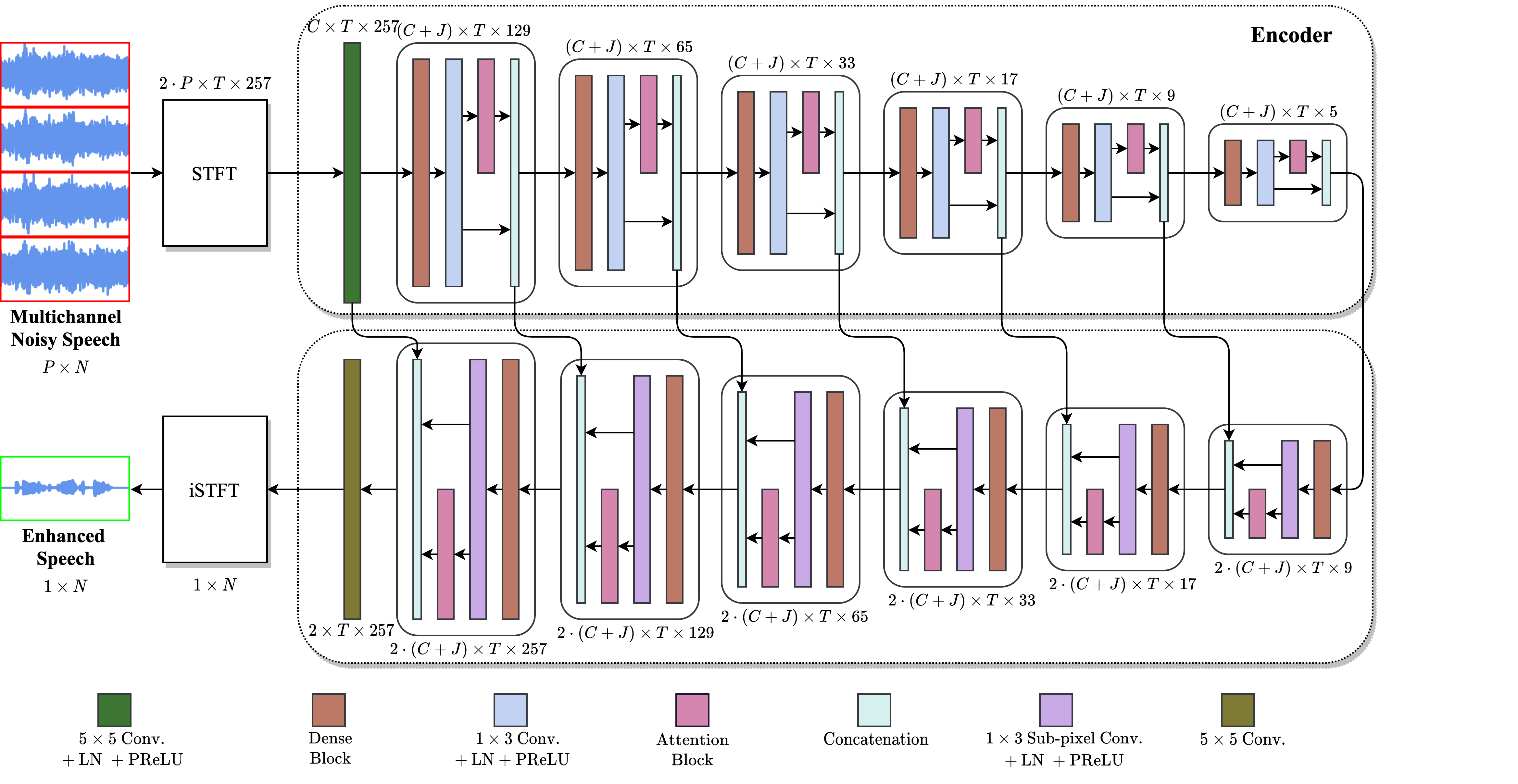}
\caption{The proposed ADCN for multi-channel complex spectral mapping.}
\label{fig:rnnt}
\end{figure*}

Next, we evaluate different models in a two-stage scheme with and without an MVDR beamformer. We empirically study two complex spectral mapping models: ADCN and DCRN from~\cite{wang2020multi}, and one waveform mapping model triple-path attentive recurrent network (TPARN) recently proposed  in ~\cite{pandey2022tparn}.

Our experimental results indicate that an MVDR beamformer becomes redundant when two different approaches are used in two stages.  Also, we obtaine significantly better results when a stronger model in the first stage is followed by a relatively weaker model in the second stage. 

\begin{comment} We find that the improvements due to the MVDR beamformer are observed only when single-domain models, such as CSM, are used in both the stages. This is in agreement with~\cite{wang2020multi}, where authors report significant improvements from an MVDR beamformer with DCRN for CSM. We also observe that the advantage from the beamformer diminishes with a very strong DNN model in the first stage. For instance, with stronger models, such as ADCN and TPARN, we do not observe consistent or significant improvements by including the beamformer. 

Finally, we report best results by using ADCN for CSM in the first stage and TPARN for WM in the second stage. Also, with a pair of DCRN and TPARN, in which TPARN is the stronger model, best results are obtained by using TPARN in the first stage and DCRN in the second stage. In other words, for a given pair of CSM and WM model, best results are obtained by using the stronger model in the first stage. \end{comment}

%\section{Model Description}
\section{Problem Definition}
A multi-channel noisy speech \scalebox{0.94}{$\bm{x} = [\bm{x}_{1}, \dots, \bm{x}_{P}]\in \mathbb{R}^{P \times N}$} with $N$ samples and $P$ microphones is modeled as

\begin{equation}
\begin{adjustbox}{width=0.74\columnwidth}
$
\begin{split}
x_{p}(n) &= y_{p}(n) + z_{p}(n) \\
               &= h_{p}(n) \ast s(n) + z_{p}(n) \\
               &= (h_{p}^{d}(n)  + h_{p}^{r}(n)) \ast s(n) + z_{p}(n)] \\
               &= h_{p}^{d}(n) \ast s(n) + [h_{p}^{r}(n) \ast s(n) + z_{p}(n)] \\
               & = d_{p}(n) + [r_{p}(n) + z_{p}(n)] \\
               &= d_{p}(n) + u_{p}(n)
\end{split} 
$
\end{adjustbox}
\end{equation}
where $p = 1, 2, \dots, P, \ n = 0 , 1, \dots N-1$.  $\bm{s}$ is the source speech, $\bm{y}_{p}$ and $\bm{z}_{p}$ are respectively the reverberated speech and noise received at microphone $p$. $\ast$ denotes convolution operator and $\bm{h}$ is the room impulse response (RIR) of source speech. $\bm{h}^{d}$ is the direct-path RIR and $\bm{h}^{r}$ is the reverberation-path RIR of the source speech. $\bm{u}$ is the overall interference including noise and room reverberation. A multi-channel speech enhancement algorithm aims at obtaining a good estimate $\hat{\bm{d}}_{r}$ of the direct-path speech at a reference microphone $r$ from multi-channel noisy recording $\bm{x}$.

\section{Attentive Dense Convolutional Network}
The architecture of the proposed ADCN is shown in Fig. 1. It is a UNet architecture with an encoder and a decoder. The input to ADCN, $\bm{X} = \text{STFT}(\bm{x})$, is of shape $2\cdot P \times T \times 257$ with $T$ frames. It is transformed to shape $C \times T \times 257$ using a $5 \times 5$ convolutional layer with layer normalization (LN) and parametric ReLU (PReLU). Next, it is processed using a stack of $6$ encoder blocks and $6$ decoder blocks. The output of a decoder block is concatenated with the output from a corresponding symmetric block in the encoder. The final output is computed by a $5 \times 5$ convolutional layer with $2$ output channels. The output waveform is obtained using an inverse STFT (iSTFT) layer at the output.

The encoder block comprises a stack of a dense block, a $1 \times 3 $ convolutional block using a stride of $2$ for downsampling with LN and PReLU, and an attention block. The output of the attention block is concatenated with its input to get the final output. The decoder block is similar to the encoder block except that it uses $1 \times 3 $ sub-pixel convolution for upsampling \cite{pandey2021dense} in the place of strided convolution for downsampling. 

\begin{figure}[!b]
\centering
\includegraphics[width=0.9\columnwidth, keepaspectratio]{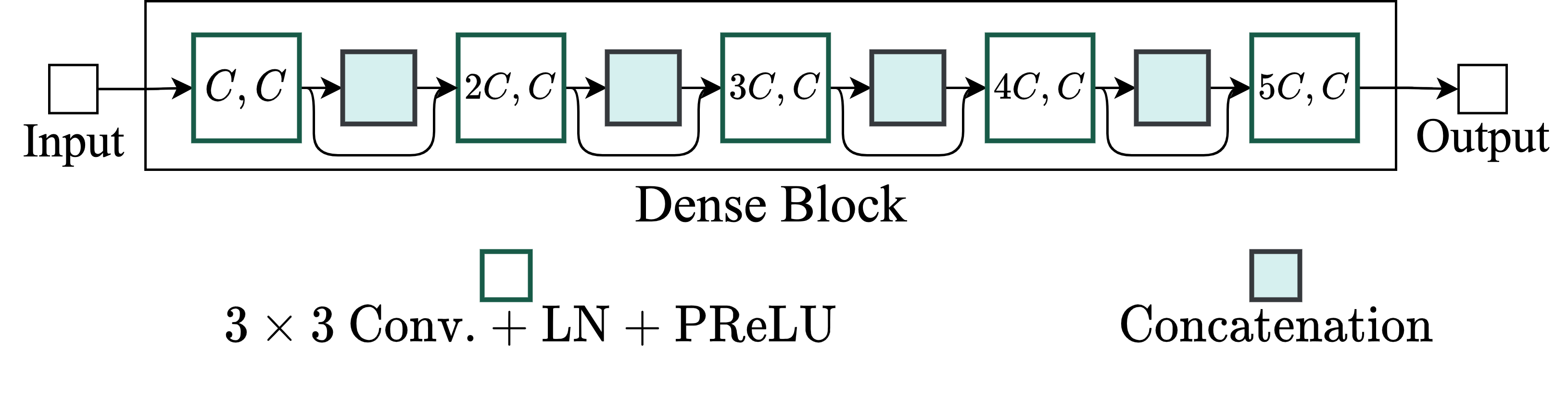}
\caption{Dense block in ADCN. $a$ and $b$ in $a, b$ inside a box respectively denote the number of input and output channels.}
\label{fig:rnnt}
\end{figure}

The architecture of the dense block is shown in Fig. 2. It comprises a stack of five $3 \times 3$ convolutional layer with $C$ output channels, LN and PReLU. The input to a given convolutional layer in a dense block is a concatenation of the block input and outputs from preceding convolutional layers in the block. 

The architecture of the attention block is shown in Fig. 3. An input of shape $C \times T \times L$ is first transformed using three separate $1 \times 1$ convolutional layers to get query $\bm{Q}$, key  $\bm{K}$, and value $\bm{V}$ of shapes $E \times T \times L$, $E \times T \times L$, and $J \times T \times L$ respectively and then rearranged  to 2d tensors of shapes $ T \times E \cdot L$,  $T \times E \cdot L$, and $T \times J \cdot L $. Next, the output from attention is computed as $\bm{A} = \text{Softmax}(\bm{Q}{\bm{K}^{T}})\bm{V}$. Finally,  $\bm{A}$ is rearranged to a 3d tensor of shape  $J \times T \times L$. 

\begin{figure}[h]
\centering
\includegraphics[width=0.96\columnwidth, keepaspectratio]{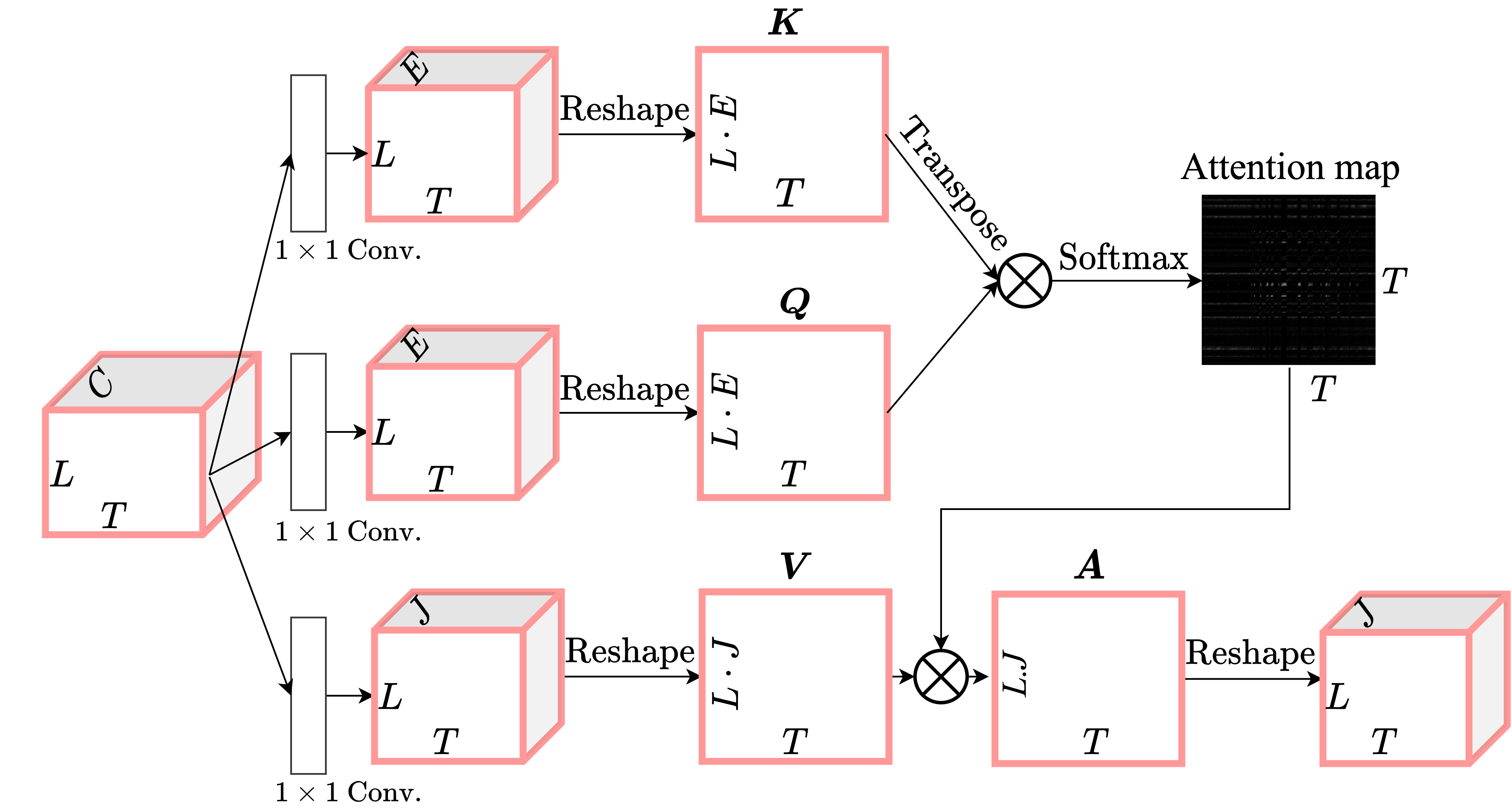}
\caption{Attention block in ADCN.}
\label{fig:rnnt}
\end{figure}

\begin{figure}[!b]
\label{fig_pcm}
\centering
\begin{subfigure}[!b]{0.5\textwidth}
\centering
\includegraphics[width=0.96\textwidth]{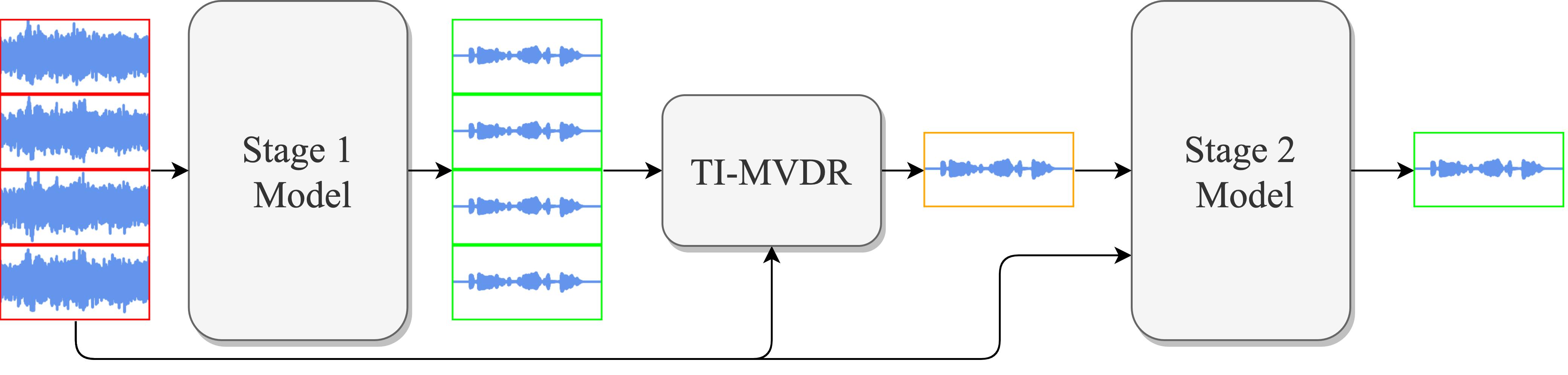}
\caption{The two-stage approach with a beamformer.}
\end{subfigure}\par 
\bigskip
\begin{subfigure}[h]{0.5\textwidth}
\centering
\includegraphics[width=0.68\textwidth]{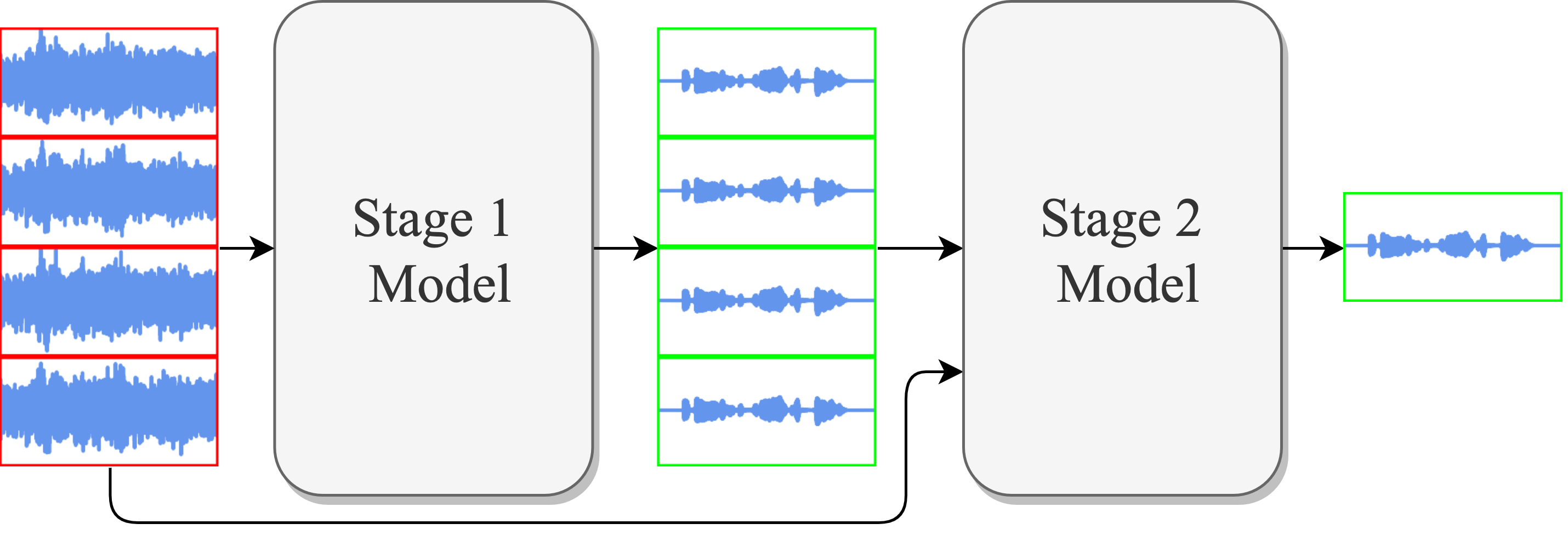}
\caption{The two-stage approach without a beamformer.}
\end{subfigure}
\caption{}
\label{fig_re_im}
\end{figure}

\section{Two-Stage Multichannel Speech Enhancement}
We evaluate three models: DCRN, ADCN and TPARN with the following approaches to two-stage processing. 

\subsection{Two-stage Approach with a Beamformer}
The two-stage approach with an MVDR beamformer is shown in Fig.~\ref{fig_re_im} (a). First, a DNN is trained to estimate enhanced speech at all channels. TPARN can output enhanced signals at all channels simultaneously as it is a multiple-input and multiple-output (MIMO) model. DCRN and ADCN, on the other hand, are multiple-input and single-output (MISO) model, and hence, require enhancing all channels independently by running the enhancement model $P$ times for $P$ channels. The output for the $m^{th}$ microphone is computed by using a circularly shifted input $[\bm{x}_{m}, \bm{x}_{m+1}, \dots,  \bm{x}_{m-2}, \bm{x}_{m -1}]$ \cite{wang2020multi}. This strategy works because we use a symmetric circular microphone array.  

 Next, enhanced speech at all channels are used to estimate the coefficients of a time-invariant MVDR (TI-MVDR) beamformer using following equations. 
\begin{equation}
\begin{split}
\hat{\Phi}^{(d)}(f) &= \frac{1}{T} \sum_{t=1}^{T}\hat{D}(t, f)\hat{D}(t, f)^{H}\\
\hat{\Phi}^{(u)}(f) &= \frac{1}{T} \sum_{t=1}^{T}\hat{U}(t, f)\hat{U}(t, f)^{H}\\
\end{split}
\end{equation}
where $\bm{\hat{D}} = \text{STFT}(\bm{\hat{d}}) \in \mathbb{C}^{P \times T \times F}$ has $T$ frames and $F$ frequency bins, $\bm{\hat{U}} = \bm{X} - \bm{\hat{U}}$, and $\bm{\hat{D}}(t, f)$ is the value at frame $t$ and frequency bin $f$.

In our experiments, sound sources are assumed to be static within each utterance, therefore, time-invariant MVDR (TI-MVDR) is a better choice than time-varying beamformer \cite{wang2021multi}. The relative transfer function with respect to the reference microphone is computed as
\begin{align}
\bm{\hat{c}}_{r}(f)  = \mathbb{P}\{ \hat{\Phi}^{d}(f) \} / \mathbb{P}\{ \hat{\Phi}^{d}_{r}(f) \}
\end{align}

where $\mathbb{P}$ extracts the principal eigenvector and $\hat{\Phi}^{d}_{r}(f)$ is the $r^{th}$ component of $\hat{\Phi}^{d}(f)$. The MVDR beamformer is computed as 
\begin{equation}
\bm{\hat{w}}_{r}(f) = \frac{\bm{\hat{\Phi}}^{u}(f)^{-1}\bm{\hat{c}}_{r}(f)}{\bm{\hat{c}}_{r}(f)^{H}\bm{\hat{\Phi}}^{u}(f)^{-1}\bm{\hat{c}}_{r}(f)}
\end{equation}
The beamformer output is computed as 
\begin{equation}
\hat{BF}_{r}(t, f) = \bm{\hat{w}}_{r}(f)^{H}\bm{X}(t, f)
\end{equation}
A waveform from the beamformer output is obtained  as
\begin{equation}
\bm{\hat{b}}_{r} = \text{iSTFT}(\bm{\hat{BF_{r}}})
\end{equation} 
Finally, a second DNN model is trained to map the speech from beamformer and the noisy multi-channel speech to enhanced speech. The input to DCRN and ADCN is a concatenation of $\bm{\hat{b}}_{r}$ and $\bm{x}$ along the channel dimension in STFT. The input to TPARN is a concatenation of $\bm{\hat{b}}$ and $\bm{x}$ along the frame dimension as TPARN requires a sequential input across channels \cite{pandey2022tparn, pandey2022tadrn}. 

\subsection{Two-stage Approach without a Beamformer}
The two-stage approach without a beamformer is shown in Fig.~\ref{fig_re_im} (b). In this approach, a DNN is trained first to get an estimate of enhanced speech at all channels and then an another DNN is trained to map enhanced speech and noisy speech at all channels to the enhanced speech at the reference channel. The input to DCRN and ADCN is a concatenation of $\bm{\hat{d}}$ and $\bm{x}$ along the channel dimension in STFT. The input to TPARN is a concatenation of $\bm{\hat{d}}$ and $\bm{x}$ along the frame dimension.

\begin{table}[t!]
\centering
\caption{Model comparisons for single-stage multi-channel speech enhancement.}
\begin{adjustbox}{width=0.7\columnwidth}
\begin{tabular}{|c|c|c|c|c|c|c|}
\hline
Test Dataset & \multicolumn{3}{c|}{WSJCAM0} & \multicolumn{3}{c|}{DNS} \\
\hline
Test Metric & SI-SDR & STOI & PESQ & SI-SDR & STOI & PESQ \\
\hline
\hline
Unprocessed & -3.8 & 70.9 & 1.63 & -7.6 & 63.8 & 1.38 \\
\hline
\hline
DCRN & 9.4 & 96.5 & 3.31 & 4.6 & 90.1 & 2.57 \\
TPARN & 10.4 & 96.9 & \textbf{3.43} & \textbf{8.4} & 91.9 & 2.75 \\
ADCN & \textbf{12.0} & \textbf{97.3} & 3.42 & 7.8 & \textbf{92.3} & \textbf{2.84} \\
\hline
\hline
\end{tabular}
\end{adjustbox}
\end{table}
\begin{table}[t!]
\centering
\caption{Model comparisons for two-stage multi-channel speech enhancement with and without beamforming.}
\begin{adjustbox}{width=0.94\columnwidth}
\begin{tabular}{|c|c|c||c|c|c||c|c|c|}
\hline
\multicolumn{3}{|c||}{Test Dataset} & \multicolumn{3}{c||}{WSJCAM0} & \multicolumn{3}{c|}{DNS} \\
\hline
Stage1$\downarrow$ & Stage2 $\downarrow$ & Type $\downarrow$ & SI-SDR & \ \ STOI \ \  & \ \ PESQ \ \ & SI-SDR & \ \ STOI \ \   & \ \ PESQ \ \  \\
\hline
\hline
\multicolumn{2}{|c|}{Unprocessed} & \xmark & -3.8 & 70.9 & 1.63 & -7.6 & 63.8 & 1.38 \\
\hline
\hline
%\multirow{7}{*}{ DCRN } & \xmark & (a) & 9.4 & 96.5 & 3.31 & 4.6 & 90.1 & 2.57 \\
%\cline{2-9}
\multirow{6}{*}{ DCRN } & \multirow{2}{*}{ DCRN } & \textbf{(a)} & 10.7 & 97.1 & 3.43 & 5.6 & 91.6 & 2.69 \\
& & (b) & 9.9 & 96.8 & 3.46 & 6.9 & 91.0 & 2.64 \\
\cline{2-9}
& \multirow{2}{*}{ TPARN } & (a) & 11.4 & 97.2 & 3.51 & 7.3 & 90.9 & 2.63 \\
& & \textbf{(b)} & 11.1 & 97.3 & 3.56 & 8.3 & 92.2 & 2.80 \\
\cline{2-9}
& \multirow{2}{*}{ ADCN } & \textbf{(a)} & 12.5 & 97.4 & 3.44 & 8.0 & 93.0 & 2.87 \\
& & (b) & 11.2 & 97.1 & 3.47 & 7.5 & 91.5 & 2.73 \\
\hline
\hline
%\multirow{7}{*}{ TPARN } & \xmark & (a) & 10.4 & 96.9 & 3.43 & 8.4 & 91.9 & 2.75 \\
%\cline{2-9}
\multirow{6}{*}{ TPARN } & \multirow{2}{*}{ DCRN } & (a) & 11.2 & 97.2 & 3.45 & 6.7 & 92.0 & 2.73 \\
& & \textbf{(b)} & 12.3 & 97.5 & 3.55 & 9.2 & 93.0 & 2.85 \\
\cline{2-9}
& \multirow{2}{*}{ TPARN } & \textbf{(a)} & 11.3 & 97.2 & 3.52 & 8.1 & 91.8 & 2.71 \\
& & \textbf{(b)} & 12.1 & 96.9 & 3.47 & 8.5 & 92.0 & 2.76 \\
\cline{2-9}
& \multirow{2}{*}{ ADCN } & \textbf{(a)} & 12.9 & 97.4 & 3.42 & 8.9 & 93.5 & 2.95 \\
& & \textbf{(b)} & 12.3 & 97.5 & 3.51 & 9.6 & 93.2 & 2.92 \\
\hline
\hline
%\multirow{7}{*}{ ADCN } & \xmark & (a) & 12.0 & 97.3 & 3.42 & 7.8 & 92.3 & 2.84 \\
%\cline{2-9}
\multirow{6}{*}{ ADCN }& \multirow{2}{*}{ DCRN } & (a) & 10.9 & 97.0 & 3.43 & 6.6 & 92.0 & 2.75 \\
& & \textbf{(b)} & 12.7 & 97.5 & 3.47 & 8.6 & 92.9 & 2.85 \\
\cline{2-9}
& \multirow{2}{*}{ TPARN } & (a) & 11.8 & 97.2 & 3.50 & 7.9 & 91.4 & 2.67 \\
& & \textbf{(b)} & \textbf{13.8} & \textbf{98.0} & \textbf{3.64} & \textbf{10.0} & \textbf{93.7} & \textbf{2.99} \\
\cline{2-9}
& \multirow{2}{*}{ ADCN } & \textbf{(a)} & 12.7 & 97.5 & 3.47 & 8.5 & 93.4 & 2.93 \\
& & \textbf{(b)} & 12.4 & 97.5 & 3.48 & 8.9 & 92.9 & 2.89 \\
\hline
\end{tabular}
\end{adjustbox}
\end{table}
  
\section{Experiments}
\subsection{Experimental Settings}
We use a four-microphone circular array of radius of $10$ cm with equal spacing between microphones. All the models are trained and evaluated on two different datasets. The first dataset is created using speakers from the WSJCAM0 \cite{robinson1995wsjcamo} dataset and noises from the REVERB challenge \cite{kinoshita2016summary}. A uniform T60 from $[0.2, 1.2]$ seconds is used for reverberation and a uniform SNR from $[5, 20]$ dB is used for noise. An algorithm for generating this dataset is given in \cite{wang2020multi}. The second dataset is created from the DNS 2020 corpus\footnote{\footnotesize{\url{https://github.com/microsoft/DNS-Challenge/blob/master/LICENSE}}}~\cite{reddy2020interspeech}. For this dataset, T60 is used from $[0.2, 1.2]$ seconds and SNR is used from $[-10, 10]$ dB. The DNS dataset can be considered more challenging than the WSJ0CAM / REVERB dataset as it uses diverse and difficult non-stationary noises with low SNR values. The data generation algorithm for the DNS dataset is given in \cite{pandey2022tparn}.
\begin{comment}
\setlength{\textfloatsep}{0.4cm}
\begin{algorithm}[!t]
\caption{Ad-hoc array dataset spatialization process.}
\begin{algorithmic}
\footnotesize
\For {\emph{split} in \{train, test, validation \}}
\For {speech utterances in \emph{split} }
\begin{itemize}
\item Draw room length and width from [5,10] m, and height from [3, 4] m;
\item Draw 6 microphone locations inside the room
\item Draw 1 speech source location inside the room;
\item Draw $N_{ns}$, number of noise sources, from [5, 10]
\item Draw $N_{ns}$ noise locations inside  the room
\item Generate RIRs corresponding to speech source location and $N_{ns}$ noise locations for all microphone locations
\item Draw $N_{ns}$ noise utterances from noises in \emph{split} 
\item Propagate speech and noise signals to all mics by convolving with corresponding RIRs
\item Draw a value $snr$ from [-10, 10] dB, and add speech and noises at each mic using a scale so that the overall direct speech SNR is $snr$;
\end{itemize}
\EndFor
\EndFor
\end{algorithmic}
\end{algorithm} 
\setlength{\floatsep}{0.4cm}
\end{comment}
%\subsection{Experimental settings}

All the utterances are resampled to 16 kHz. We use a frame size of $32$ ms, frame shift of $8$ ms, $C = 64, E=5$, and $J=32$ for ADCN. TPARN and DCRN are trained using methods proposed in their original studies. ADCN is trained using a phase constrained magnitude (PCM) loss \cite{pandey2021dense}. All the models are trained for 100 epochs with a batch size of $8$ $4$-s long utterances randomly extracted at the training time. The initial learning rate is set to $0.0004$ and is scaled by half if the validation score does not improve for five consecutive epochs.   

The first microphone, $r=1$, is used for objective evaluation. All the models are evaluated using short-time objective intelligibility (STOI) \cite{taal2011algorithm}, perceptual evaluation of speech quality (PESQ) \cite{rix2001perceptual}, and scale-invariant signal-to-distortion ratio (SI-SDR). STOI is reported in percentage. 
% \subsection{Baselines}

%\vspace{-10pt}
%\vspace{-10pt}

\subsection{Experimental Results}
First, we compare DCRN, ADCN and TPARN in Table 1 for single-stage end-to-end training. ADCN obtains best results in all the cases except for SI-SDR at DNS where it is slightly worse that TPARN. A general performance order of these models is DCRN $<$ TPARN $<$ ADCN. Even though TPARN is worse than ADCN, it has computational advantages over ADCN and DCRN. For example, using ADCN in the first stage requires $P$ forward passes for $P$ channels, whereas TPARN can enhance signals at all channels in one pass. 

Next, we compare two approaches: (a) two-stage with a beamformer, (b) two-stage without a beamformer. We consider an overall improvement over two datasets between (a), (b) and make bold the one with better improvements. We highlight both (a) and (b) if their performances are similar. 

Firstly, we observe that with a complex spectral mapping model DCRN in the first stage, beamformer is better for complex spectral mapping models DCRN and ADCN, but worse for the waveform mapping model TPARN in the second stage. This suggest that for a weaker complex spectral mapping model, beamformer is helpful but only for models with the same approach, complex spectral mapping, in the second stage.

Next,  we observe that with TPARN in the first stage, beamformer is worse for DCRN and and similar for ADCN and TPARN in the second stage. This suggest that for a relatively stronger model TPARN, beamformer does not provide any consistent improvement.

Further, we can see that with ADCN in the first stage, beamformer obtains worse results with DCRN and TPARN and comparable results with ADCN. These comparisons indicate that the beamformer is helpful only if the first stage model is weak and the second stage model uses same approach as the first stage model. 

Finally, we note that the best results are reported with TPARN followed by DCRN for the pair (TPARN, DCRN) and with ADCN followed by TPARN for the pair (TPARN, ADCN). This suggests that a much better performance can be obtained without a beamformer by using different approaches, complex spectral mapping and waveform mapping in two stages and employing the stronger model in the first stage. Also, we find that ADCN followed by TPARN obtains significantly better results than the second best.  
\section{Conclusions}
We have proposed a novel attentive dense convolutional network for multi-channel speech enhancement. We have also proposed a two-stage approach that obtains excellent results without a beamformer. The proposed approach uses ADCN for complex spectral mapping in the first stage and TPARN for waveform mapping in the second stage. Future research includes evaluating the effectiveness of the proposed approach for ASR improvements.

%\bibliographystyle{IEEEbib}
%\bibliography{mybib}

%biblatex
\setlength\bibitemsep{0.5em}
\atColsBreak{\vskip0.5em}
\printbibliography

\end{document}